\documentclass[nofootinbib,twocolumn,preprintnumbers]{revtex4-1}
\usepackage{amsmath}
\usepackage{amssymb}
\usepackage{graphicx}
\usepackage{booktabs}

\newcommand{\order}[1]{{\cal O}\left(#1\right)}

\newcommand{\as}{\alpha_s}

\newcommand{\pb}{\;\mathrm{pb}}

\newcommand{\GeV}{\;\mathrm{GeV}}
\newcommand{\TeV}{\;\mathrm{TeV}}
\newcommand{\ptjv}{p_{\rm t,veto}}
\newcommand{\cF}{{\cal F}}
\newcommand{\NNLL}{\text{NNLL}}

\begin{document}

\title{Higgs and Z-boson production with a jet veto
}

\preprint{CERN-PH-TH/2012-173}
\preprint{LPN12-061}
\preprint{OUTP-12-14P}
\preprint{ZU-ZH 10/12}

\author{Andrea Banfi,$^1$ Pier Francesco Monni,$^2$ Gavin P. Salam,$^{3,4,5}$ Giulia
  Zanderighi\,$^6$\vspace{1em}}

\affiliation{$^1$ Albert-Ludwigs-Universit\"at Freiburg, 
   Physikalisches Institut, D-79104 Freiburg, Germany \\
$^2$ Institut f\"ur Theoretische Physik, Universit\"at Z\"urich, CH-8057 Z\"urich, Switzerland\\
$^3$ CERN, PH-TH, CH-1211 Geneva 23, Switzerland\\
$^4$ Department of Physics, Princeton University, Princeton, NJ 08544, USA\\
$^5$ LPTHE; CNRS UMR 7589; UPMC Univ.\ Paris 6; Paris, France\\
$^6$ Rudolf Peierls Centre for Theoretical Physics,
  1 Keble Road, University of Oxford, UK}

\begin{abstract}
  We derive first next-to-next-to-leading logarithmic
  resummations for jet-veto efficiencies in Higgs and Z-boson
  production at hadron colliders.
  Matching with next-to-next-to-leading order results allows us
  to provide a range of phenomenological predictions for the LHC, including
  cross-section results, detailed uncertainty estimates and
  comparisons to current widely-used tools.
\end{abstract}

\pacs{13.87.Ce,  13.87.Fh, 13.65.+i}

\maketitle

In searches for new physics at hadron colliders such
as the Tevatron and CERN's Large Hadron Collider (LHC), in order to
select signal events and reduce backgrounds, events are often
classified according to the number of hadronic jets --- collimated
bunches of energetic hadrons --- in the final state.
A classic example is the search for Higgs production via gluon
fusion with a subsequent decay to
$W^+W^-$~\cite{Chatrchyan:2012ty,ATLAS:2012sc}. 
A severe background comes from $t\bar t$ production, whose
decay products also include a $W^+W^-$ pair. 
However, this background can be separated from the signal because its
$W^+W^-$ pair usually comes together with hard jets, since in each top
decay the $W$ is accompanied by an energetic ($b$) quark.

Relative to classifications based on objects such as leptons (used
e.g.\ to identify the $W$ decays), one of the difficulties of 
hadronic jets is that they may originate not just from the decay of a
heavy particle, but also as Quantum Chromodynamic (QCD) radiation.
This is the case in our example, where the incoming gluons that fuse
to produce the Higgs quite often radiate additional partons.
Consequently, while vetoing the presence of jets eliminates much of the
$t\bar t$ background, it also removes some fraction of signal
events. 
To fully interpret the search results, including measuring Higgs
couplings, it is crucial to be able to predict the fraction of the
signal that survives the jet veto, which depends for example on the
transverse momentum threshold $\ptjv$ used to identify vetoed jets.

One way to evaluate jet-veto efficiencies is to use a fixed-order
perturbative expansion in the strong coupling $\as$, notably to
next-next-to-leading order (NNLO), as in the Higgs-boson production
calculations of~\cite{Catani:2001cr,Anastasiou:2005qj,Grazzini:2008tf}.
Such calculations however become unreliable for $\ptjv \ll M$, with
$M$ the boson mass, since large terms $\as^n L^{2n}$ appear
($L=\ln(M/\ptjv)$) in the cross-section to all orders in the coupling
constant.
These enhanced classes of terms can, however, be resummed to all
orders in the coupling, often involving a functional form
$\exp(L g_1(\as L) + g_2(\as L) + \as
g_3(\as L)/\pi + \ldots)$.

There exist next-to-next-to-leading logarithmic (NNLL) resummations,
involving the $g_n(\as L)$ functions up to and including $g_3$, for a
number of quantities that are more inclusive than a jet veto: e.g. a
Higgs or vector-boson transverse
momentum~\cite{Bozzi:2003jy,Bozzi:2005wk,Bozzi:2010xn,Becher:2010tm},
the beam thrust \cite{Berger:2010xi}, and related
observables~\cite{Banfi:2011dx,Banfi:2012du}.
To obtain estimates for jet vetoes, some of these calculations have
been compared to or used to 
reweight~\cite{Davatz:2006ut,Anastasiou:2008ik,Anastasiou:2009bt,Berger:2010xi,Dittmaier:2012vm}
parton-shower predictions~\cite{Sjostrand:2006za,Corcella:2002jc}
matched to NLO results~\cite{Frixione:2002ik,Nason:2004rx}.
However, with reweighting, neither the NNLO nor NNLL accuracy of the
original calculation carry through to the jet veto prediction.

Recently there has been progress towards NNLL calculations of the jet
veto efficiency itself.
Full NLL results and some NNLL ingredients for Higgs and
vector-boson were provided in~\cite{BSZ12}.
Ref.~\cite{Becher:2012qa} used these and other ingredients in the
soft-collinear effective theory framework to consider resummation for
the Higgs-boson case beyond NLL accuracy.
In this letter we show how to use the results of \cite{BSZ12}
together with those from boson $p_t$
resummations~\cite{Bozzi:2003jy,Bozzi:2005wk,Bozzi:2010xn,Becher:2010tm}
to obtain full NNLL accuracy.
We also examine the phenomenological impact of our results, including
a matching to NNLO predictions.
Given the ubiquity of jet cuts in hadron-collider analyses, the
understanding gained from our analysis has a potentially wide range of
applications.

The core of boson transverse-momentum ($p_t^B$) resummations lies in
the fact that soft, collinear emissions at disparate rapidities are
effectively emitted independently. Summing over all independent
emissions, one obtains the differential boson $p_t$ cross section
\begin{equation}
  \label{eq:ptB}
  \frac{d\Sigma^{(B)}}{d^2p_t^B} \!=\! \sigma_0 \!\!\!\int\! \!\frac{d^2 b}{4\pi^2} e^{-i b.p_t^B} 
  \!\sum_{n} \!\frac{1}{n!} \prod_{i=1}^n\! \!\int\![dk_i] M^2(k_i)
  (\!e^{i b. k_{ti}}-1\!),
\end{equation}
where $\sigma_0$ is the leading-order total cross section, $[dk_i]
M^2(k_i)$ is the phase-space and matrix-element for emitting a soft,
collinear gluon of momentum $k_i$, while the exponential factors and
$b$ integral encode in a factorised form the constraint relating
the boson $p_t$ and those of individual emissions
$\delta^2(p_t^B - \sum_{i=1}^n k_{ti})$ \cite{Parisi:1979se}.
The $-1$ term in the round brackets arises because, by unitarity,
virtual corrections come with a weight opposite to that of real
emissions, but don't contribute to the $p_t^B$ sum.

To relate Eq.~(\ref{eq:ptB}) to a cross section with a jet-veto, let
us first make two simplifying assumptions: that the
independent-emission picture is exact and that a jet algorithm
clusters each emission into a separate jet. The resummation for the
cross section for the highest jet $p_t$ to be below some
threshold $p_t^J$, considering jets at all rapidities,  is then
equivalent to requiring all emissions to be below that threshold:
\begin{align}
  \label{eq:ptJ-start}
  \Sigma^{(J)}(p_t^J) &= \sigma_0 \sum_{n=0}^\infty \frac{1}{n!} 
  \prod_{i=1}^n \int [dk_i] M^2(k_i) (\Theta(p_t^J - k_{ti}) - 1)
  \nonumber
  \\
  &=  \sigma_0 \exp \left[-\int [dk_i] M^2(k_i) \Theta(k_{ti} - p_t^J)
  \right]\,,
\end{align}
with the same universal 
matrix element $M^2(k_i)$ entering Eqs.~(\ref{eq:ptB}) and
(\ref{eq:ptJ-start}).

Eq.~(\ref{eq:ptJ-start}) is clearly an oversimplification.
Firstly, even within the independent emission picture, two emissions
close in rapidity $y$ and azimuth $\phi$ can be clustered
together into a single jet.
Let us introduce a function $J(k_1,k_2)$ that is $1$ if $k_1$ and
$k_2$ are clustered together and $0$ otherwise.
Concentrating on the $2$-emission contribution to
Eq.~(\ref{eq:ptJ-start}), one sees that clustering
leads to a correction given by the difference between the veto with
and without clustering:
\begin{multline}
  \label{eq:ptJ-clustering}
  \cF^{\text{clust}} \sigma_0  = \frac{\sigma_0}{2!}\int
  [dk_1][dk_2] M^2(k_1)M^2(k_2)\,
  \times 
  J(k_1,k_2) 
  \\
  (\Theta(p_t^J - k_{t,12}) - \Theta(p_t^J - k_{1,t})\Theta(p_t^J -
  k_{2,t}))\,.
\end{multline}
where $k_{12} = k_1 + k_2$ (throughout, we assume standard $E$-scheme
recombination, which adds 4-vectors).
This contribution has a logarithmic structure $\as^2L$, i.e.\ NNLL,
with each emission leading to a power of $\as$, while the $L$ factor
comes from the integral over allowed rapidities ($|y|\lesssim
\ln(M/\ptjv)$).

For more than two emissions, two situations are possible: (1) three or
more emissions are close in rapidity, giving extra powers of $\as$
without extra log-enhancements (N$^3$LL and beyond); (2) any number of
extra emissions are far in rapidity, each giving a factor $\as L$,
i.e. also NNLL. The latter contribution is simple because,
independently of whether the two nearby emissions clustered, those
that are far away must still have $p_{ti} < \ptjv$. Thus the full
``clustering'' correction to the independent-emission picture is a
multiplicative factor $(1 + \cF^{\text{clust}})$, as first derived in
detail in the appendix of~\cite{BSZ12} using results from
\cite{caesar}.

For the generalised-$k_t$ jet-algorithm
family~\cite{Catani:1993hr,Kt-EllisSoper,Cam,Aachen,Cacciari:2008gp},
with a jet radius parameter $R$, we have
$J(k_1,k_2) = \Theta(R^2 - (y_1-y_2)^2- (\phi_1-\phi_2)^2)$. At
NNLL accuracy Eq.~(\ref{eq:ptJ-clustering}) evaluates to
$\cF^{\text{clust}} = 4
\as^2(\ptjv) C L f^\text{clust}(R)/\pi^2$  with~\cite{BSZ12}
\begin{align}
  \label{eq:Findep}
  f^\text{clust}(R) = \left(-\frac{\pi^2 R^2}{12} + \frac{R^4}{16}\right)C\,,
\end{align}
for $R < \pi$; $C$ is $C_F=\frac43$ or $C_A=3$ respectively for incoming
quarks (e.g.\ $q\bar q \to Z$) or incoming gluons (e.g.\ $gg \to H$). 

Next, we address the issue that gluons are not all emitted
independently. 
This is accounted for in Eq.~(\ref{eq:ptB}) because, to order
$\as^2$, the $M^2(k)$ quantity that appears there should be
understood as an effective matrix element
\begin{multline}
  \label{eq:Meffective}
  [dk] M^2(k) 
  = [dk] \big(M^2_{1}(k) + M^2_{\text{1-loop}}(k)\big) 
  \\ + \int d^2k_t [dk_a][dk_b] M^2_{\text{correl}}(k_a, k_b) 
  \delta^2(k_{t,ab} - k_t)\,
\end{multline}
where $M^2_{1}(k)$ is the pure $\order{\as}$ matrix element,
$M^2_{\text{correl}}(k_a, k_b)$ the correlated part of the matrix
element for emission of two soft-collinear gluons or a
quark-antiquark
pair, including relevant symmetry factors, 
and $M^2_{\text{1-loop}}$ the corresponding part of the $\as^2$
1-loop matrix element.
The separation into correlated and independent emissions is well
defined because of the different colour factors that 
accompany them in the generic case~\cite{Berends:1988zn,
  Dokshitzer:1992ip,Campbell:1997hg,Catani:1999ss}.
The $\delta$-function in Eq.~(\ref{eq:Meffective}) extracts two-parton
configurations with the same total $p_t$ as the 1-gluon
configurations.

For a jet veto, part of the result from the effective matrix element
carries through: when two correlated emissions are clustered into a
single jet, it is their sum, ${\vec k}_{t,ab}$, that determines the
jet transverse momentum.
Therefore the same effective matrix element can be used in
Eq.~(\ref{eq:ptJ-start}), as long as one includes an additional
correction to account for configurations where the two emissions are
clustered in separate jets:
\begin{multline}
  \label{eq:ptJ-correlated}
  \!\!\!\!\cF^{\text{correl}} \sigma_0  = \sigma_0\!\!\int
  [dk_a][dk_b] M^2_{\text{correl}}(k_a,k_b)\,
  \times 
  (1 - J(k_a,k_b))
  \\
  (\Theta(p_t^J - k_{ta})\Theta(p_t^J -
  k_{tb}) - \Theta(p_t^J - k_{t,ab}) )\,.
\end{multline}
At NNLL,  $\cF^{\text{correl}} = 4
\as^2(\ptjv) C L f^\text{correl}(R)/\pi^2$  with
\begin{multline}
  \label{eq:Fcorrel}
  f^{\text{correl}}(R) =   
  \bigg(
    \frac{\left(-131+12 \pi ^2+132 \ln 2\right) C_A}{72}  
  \\
    + 
    \frac{(23-24 \ln 2) n_f}{72}
    \bigg)\ln \frac{1}{R} 
    + 0.61C_A -0.015 n_f+
    \order{R^2} 
    \,,
\end{multline}
for generalised-$k_t$ algorithms, in the limit of small
$R$. Ref.~\cite{BSZ12} includes a numerical result for all $R < 3.5$
and analytical terms up to $R^6$, used in the rest of this article.
It did not, however, make the relation with the boson
$p_t$ resummation.

All remaining contributions to a NNLL resummation, such as the 3-loop
cusp anomalous dimension or a multiplicative $C_1 \as$ term are either
purely virtual, so independent of the precise observable, or involve
at most a single real emission, so can be taken from the boson $p_t$
resummations~\cite{Bozzi:2003jy,Bozzi:2005wk,Bozzi:2010xn,Becher:2010tm}.\footnote{For
  generic processes, subtleties can arise with spin-correlation
  effects~\cite{Catani:2010pd}. These are simpler for jet vetoes,
  which don't correlate distinct collinear regions.}
Thus the full NNLL resummed cross section for the jet-veto is given
by: 
\begin{multline}
  \label{eq:SigmaNNLL-result}
  \Sigma^{(J)}_\text{\NNLL}(\ptjv) =
  \sum_{i,j}\int dx_1 dx_2 \,|{\cal M}_B|^2\delta(x_1 x_2 s - M^2)\\
  \times\bigg[f_i\!\left(x_1, e^{-L} \mu_F\right)
  f_j\!\left(x_2, e^{-L} \mu_F\right)\left(1+\frac{\alpha_{s}}{2\pi}{\cal H}^{(1)}\right) +\\
  + \frac{\alpha_{s}}{2\pi}\frac{1}{1-2\alpha_s \beta_0 L}\sum_{k}\int_{x_1}^1\frac{dz}{z}\bigg(
  C_{ki}^{(1)}(x_{1}) f_i\!\left(\frac{x_1}{z}, e^{-L} \mu_F\right)\\
  \times f_j\!\left(x_2, e^{-L} \mu_F\right) + \{(x_1,i)\,\leftrightarrow\,(x_2,j)\}\bigg)\, \bigg]\\
  (1 + \cF^{\text{clust}} + \cF^{\text{correl}})
  \times e^{L g_1(\as L) + g_2(\as L) + \frac{\as}{\pi} g_3(\as L)}\,, 
\end{multline}
where the coefficient functions ${\cal H}^{(1)}$ and $C_{ki}^{(1)}$,
and resummation functions $g_1$, $g_2$ and $g_3$ are as derived for
the boson $p_t^B$
resummation~\cite{Bozzi:2003jy,Bozzi:2005wk,Becher:2010tm} (reproduced
for completeness in the supplemental material to this
letter~\cite{BMSZadditional}, together with further discussion on the
connection to boson $p_t^B$ resummation).
The results are expressed in terms of $L = \ln({Q}/{p_{t,\text{veto}}})$, $\as
\equiv \as(\mu_R)$; the resummation, renormalisation and
factorisation scales $Q, \mu_R$ and $\mu_F$ are to be chosen of
order of $M$.

A form similar to Eq.~(\ref{eq:SigmaNNLL-result}) was derived
independently in \cite{Becher:2012qa} for Higgs production, also using
ingredients from~\cite{BSZ12}.
It differs however at NNLL in that the combination of
$f^\text{clust}+f^\text{correl}$ is accompanied by an extra
$-\zeta_3 C_A$.
Ref.~\cite{Becher:2012qa} had used a NNLL analysis of the $R \to
\infty$ limit to relate jet and boson-$p_t$ resummations.
A subtlety of this limit is that one must then account for a N$^3$LL
$\as^2 R$ term, which for $R \gtrsim \ln M/p_t$ is promoted to an
additional NNLL $\as^2 \ln M/p_t$ contribution~\cite{BMSZadditional}.

\begin{figure}[t]
  \centering
  \includegraphics[width=0.98\columnwidth]{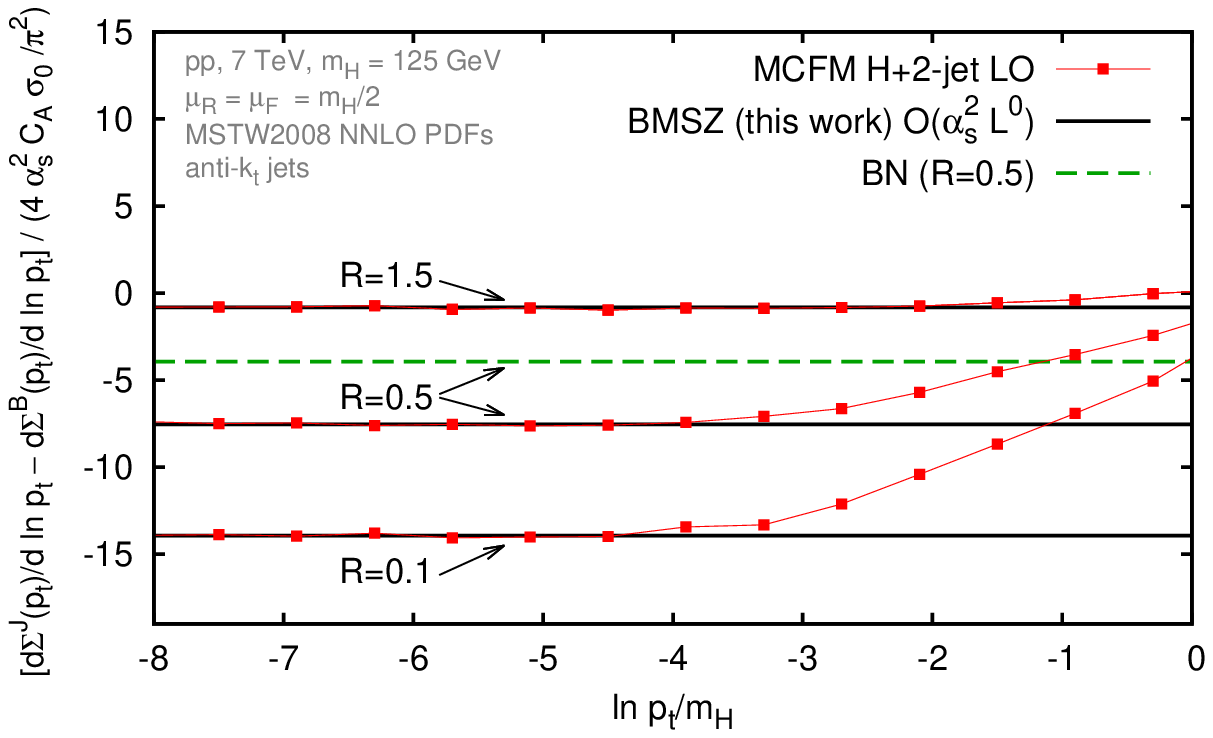}\\
  \includegraphics[width=0.98\columnwidth]{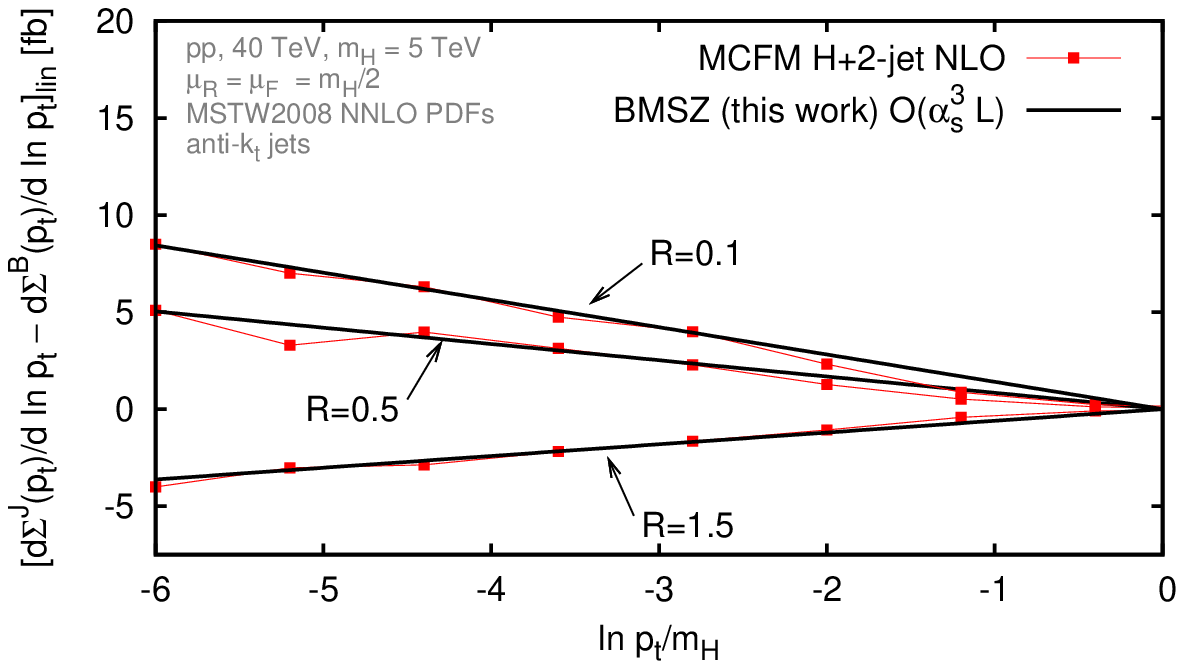}
  \caption{ %
    Upper panel: 2nd order difference between jet and
    Higgs-boson $\ln p_t$ differential distributions, showing the
    coefficient of $4\as^2 C_A \sigma_0 /\pi^2$ as determined with
    MCFM and predicted in Eq.~(\ref{eq:D2Diff}), for three $R$
    values. We also show the prediction from
    \cite{Becher:2012qa} (BN).
    Lower panel: differences at $\order{\as^3 \sigma_0}$ between jet
    and boson $\ln p_t$ differential distributions, with the expected
    $\as^3 \sigma_0 L^2$ term subtracted (denoted by a subscript
    $_\text{lin}$), showing the MCFM H+2-jet NLO result compared to
    our NNLL prediction for the $\as^3 \sigma_0 L$ term. }
  \label{fig:checks}
\end{figure}

One check of Eq.~(\ref{eq:SigmaNNLL-result}) is to expand it
in powers of $\as$, $\Sigma^{(J)}_{\NNLL}(p_t) = \as^2
\sum_{n=0}^\infty \Sigma^{(J)}_{\NNLL,n}(p_t)$, and compare
$d\Sigma^{(J)}_{\NNLL,2}(p_t)/d\ln p_t$ to the NLO Higgs+1~jet
prediction~\cite{deFlorian:1999zd,Ravindran:2002dc,Glosser:2002gm}
from MCFM~\cite{Campbell:2002tg}, $d\Sigma^{(J)}_{2}(p_t)/d\ln
p_t$. 
NNLL resummation implies control of terms $\as^2 L^3 \ldots \as^2$
(constant terms) in this quantity and so the difference between MCFM
and the 2nd order expansion of the resummation should vanish for large
$L$. This is what we find within reasonable precision.
The precision of the test can be increased if one considers the
$\order{\as^2}$ \emph{difference} between the jet and boson-$p_t$
resummations, which has fewer logarithms and so is
numerically easier to determine in MCFM. 
It is predicted to be
\begin{multline}
  \label{eq:D2Diff}
  \frac{d\Sigma^{(J)}_{\NNLL,2}(p_t)}{d\ln p_t} -
  \frac{d\Sigma^{(B)}_{\NNLL,2}(p_{t})}{d\ln p_{t}}
  = \\
  -\frac{4C \as^2 \sigma_0}{\pi^2}\left(f^\text{clust}(R) +
    f^\text{correl}(R) + \zeta_3\, C\right)\,.
\end{multline}
This is compared to MCFM's LO H+2-jet result in the upper panel of
Fig.~\ref{fig:checks}.
There is excellent agreement at small $p_t$, for each of three $R$
values.
The result of \cite{Becher:2012qa} (BN, only for $R=0.5$) is also
shown for comparison.

\begin{figure*}[t]
  \centering
  \includegraphics[width=0.48\linewidth]{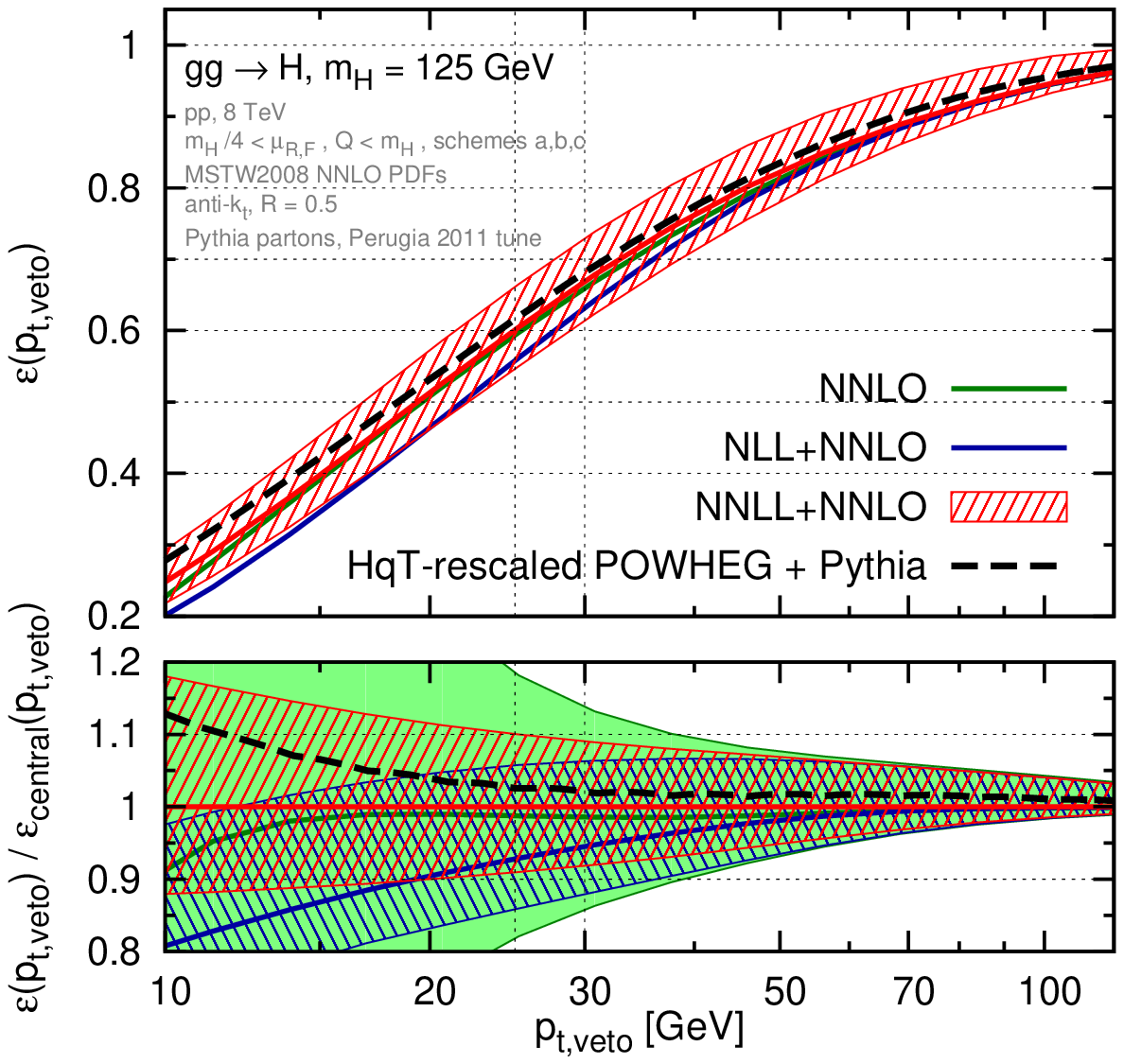}\hfill
  \includegraphics[width=0.49\linewidth]{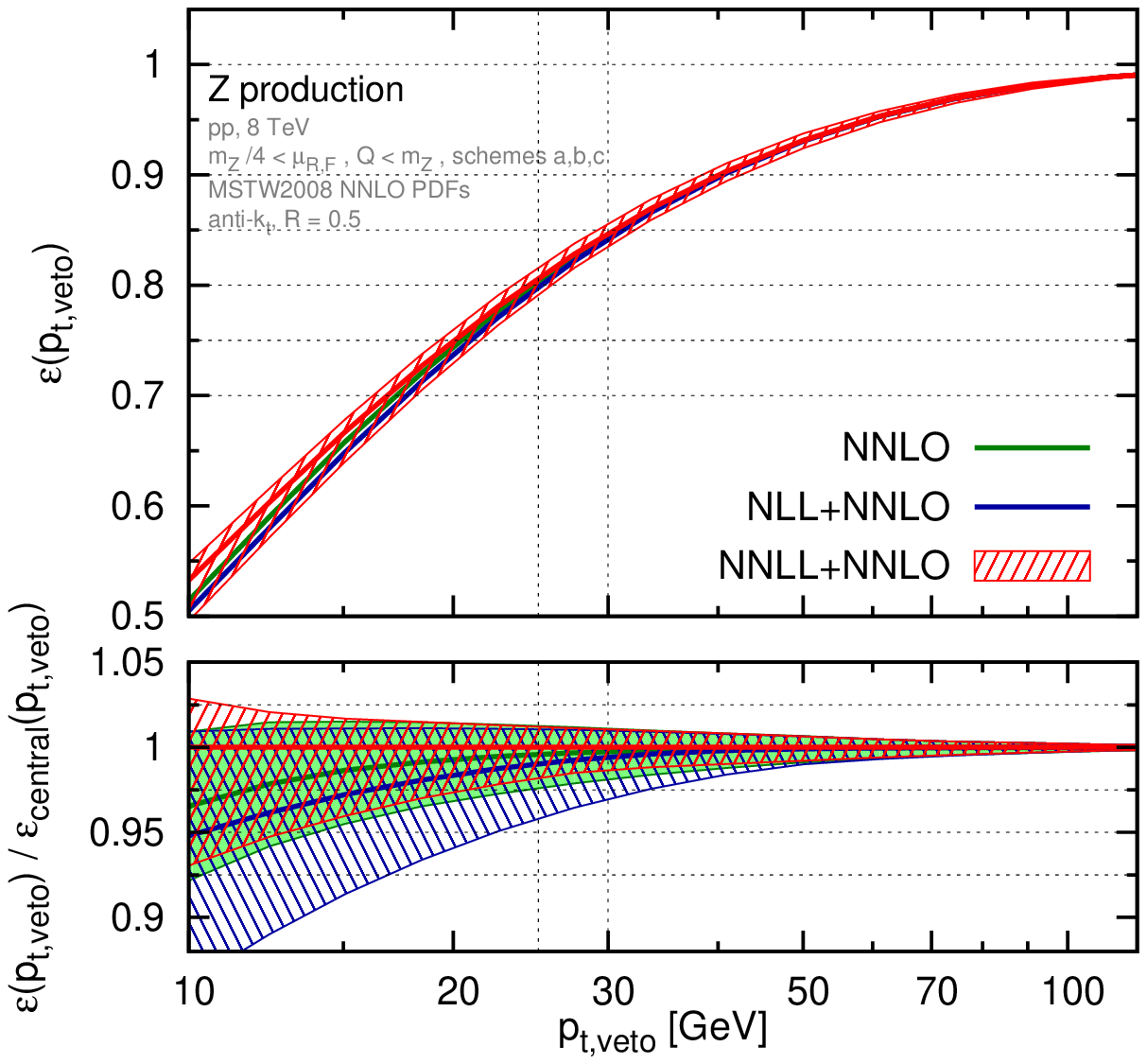}\hfill
  \caption{ %
    Comparison of NNLO, NLL+NNLO and NNLL+NNLO results for 
    jet-veto efficiencies for Higgs (left) and Z-boson (right)
    production at the 8 TeV LHC.  %
    The Higgs plot includes the result from a POWHEG (revision
    1683) \cite{Nason:2004rx,Alioli:2008tz} plus Pythia (6.426)
    \cite{Sjostrand:2006za,Skands:2010ak} simulation in which the Higgs-boson $p_t$
    distribution was reweighted to match the NNLL+NNLO prediction
    from HqT 2.0~\cite{Bozzi:2005wk} as in \cite{BSZ12}.
    The lower panels show results normalised to the central
    NNLL+NNLO efficiencies.}
  \label{fig:two-matched-v-rwgtPWG}
\end{figure*}

The above test can be extended one order further by examining the order
$\as^3 \sigma_0$ difference between the jet and boson $p_t$
differential distributions.
The comparison between our predictions and MCFM H+2-jet NLO
results~\cite{Campbell:2006xx,Campbell:2010cz} is given in
Fig.~\ref{fig:checks} (lower panel), for each of three $R$ values.
To facilitate visual interpretation of the results, the expected
$\as^3 \sigma_0 L^2$ term has been subtracted.
The residual $\as^3 \sigma_0 L$ term is clearly
visible in the MCFM results and, within the fluctuations, coincides
well with our predictions, providing a good degree of corroborating
evidence for the correctness of our results beyond order
$\as^2\sigma_0$.

To illustrate the phenomenological implications of our work, we
examine the jet veto efficiency $\epsilon(p_t) \equiv
\Sigma^{(J)}(p_t)/\sigma_\text{tot}$, where $\sigma_{tot}$ is the
total cross section, known up to
NNLO~\cite{Hamberg:1990np,Dawson:1990zj,Djouadi:1991tka,Harlander:2002wh,Anastasiou:2002yz,Ravindran:2003um}.
We combine (``match'') the resummation with fixed-order predictions,
available from fully differential NNLO boson-production
calculations~\cite{Anastasiou:2005qj,Melnikov:2006di,Catani:2009sm,Grazzini:2008tf}
or NLO boson+jet
calculations~\cite{Giele:1993dj,deFlorian:1999zd} implemented in
MCFM~\cite{Campbell:1999ah}.
We use three matching schemes, denoted $a$, $b$ and $c$,
straightforward extensions~\cite{BMSZadditional} of those used at NLL
in~\cite{BSZ12}.

Our central predictions have $\mu_R = \mu_F = Q
= M/2$ and scheme $a$ matching, with MSTW2008NNLO
PDFs~\cite{Martin:2009iq}. We use the
anti-$k_t$~\cite{Cacciari:2008gp} jet-algorithm with $R=0.5$, as
implemented in FastJet~\cite{FastJet}.
For the Higgs case we use the large $m_\text{top}$ approximation and
ignore $b\bar b$ fusion and $b$'s in the $gg\to H$ loops (corrections
beyond this approximation have a relevant
impact~\cite{Bagnaschi:2011tu,Dittmaier:2012vm}).
To determine uncertainties we vary $\mu_R$ and $\mu_F$ by a factor of
two in either direction, requiring $1/2 \le \mu_R/\mu_F \le
2$. Maintaining central $\mu_{R,F}$ values, we also vary $Q$ by a
factor of two and change to matching schemes $b$ and $c$. Our final
uncertainty band is the envelope of these variations. In the
fixed-order results, the band is just the envelope of $\mu_{R,F}$
variations.

The results for the jet-veto efficiency in Higgs and Z-boson production are
shown in Fig.~\ref{fig:two-matched-v-rwgtPWG} for $8\TeV$ LHC
collisions. 
Compared to pure NNLO results, the central value is slightly
higher and for Higgs production, the uncertainties reduced,
especially for lower $\ptjv$ values.
Compared to NNLO+NLL results~\cite{BSZ12}, the central values are
higher, sometimes close to edge of the NNLO+NLL bands;
since the NNLO+NLL results used the same approach for estimating the
uncertainties, this suggests that the approach is not unduly
conservative.
In the Higgs case, the NNLO+NNLL uncertainty band is not particularly
smaller than the NNLO+NLL one. This should not be a surprise, since
\cite{BSZ12} highlighted the existence of possible substantial
corrections beyond NNLL and beyond NNLO.
For the Higgs case, we also show a prediction from
POWHEG~\cite{Nason:2004rx,Alioli:2008tz} interfaced to Pythia
6.4~\cite{Sjostrand:2006za} at parton level (Perugia 2011 shower
tune~\cite{Skands:2010ak}), reweighted to 
describe the NNLL+NNLO Higgs-boson $p_t$ distribution from HqT
(v2.0)~\cite{Bozzi:2005wk}, as used by the LHC experiments. 
Though reweighting fails to provide NNLO or NNLL accuracy
for the jet veto, for $\ptjv$ scales of practical relevance, the
result agrees well with our central prediction.
It is however harder to reliably estimate uncertainties in
reweighting approaches than in direct calculations.

Finally, we provide central results and uncertainties for the jet-veto
efficiencies and 0-jet cross sections (in pb) with cuts (in GeV) like
those used by ATLAS and CMS, and also for a larger $R$
value:
\begin{center}
  \begin{tabular}{c|c|c|c|c|c}
    R & $\,\ptjv$\, & 
    $\epsilon^{(7 \TeV)}$ & \,$\sigma^{(7\TeV)}_\text{0-jet}$\, & 
    $\epsilon^{(8 \TeV)}$ & \,$\sigma^{(8\TeV)}_\text{0-jet}$\,
    \\[0.2em]\hline
    \phantom{x} & & & & &
    \\[-1em] 
    $0.4$\;& $25$ & \,$0.63^{+0.07}_{-0.05}$\, &
    $9.6^{+1.3}_{-1.1}$ &
    $0.61^{+0.07}_{-0.06}$ &
    $12.0^{+1.6}_{-1.4}$ 
    \\[0.4em] 
    $0.5$\;& $30$ & 
    $0.68^{+0.06}_{-0.05}$ &
    $10.4^{+1.2}_{-1.1}$ & 
    \,$0.67^{+0.06}_{-0.05}$\, &
    $13.0^{+1.5}_{-1.5}$ 
    \\[0.4em] 
    $1.0$\;& $30$ & 
    $0.64^{+0.03}_{-0.05}$ &
    $9.8^{+0.8}_{-1.1}$& 
    $0.63^{+0.04}_{-0.05}$&
    $12.2^{+1.1}_{-1.4}$
  \end{tabular}
\end{center}
Interestingly, the $R=1$ results have reduced
upper uncertainties, due perhaps to the smaller value of the NNLL $f(R)$
correction (a large $f(R)$ introduces significant $Q$-scale
dependence).
The above results are without a rapidity cut on the jets;
the rapidity cuts used by ATLAS and CMS lead only to small,
$<1\%$, differences~\cite{BSZ12}.

For the 0-jet cross sections 
above, we used total cross sections at
7~TeV and 8~TeV of $15.3^{+1.1}_{-1.2}\pb$ and $19.5^{+1.4}_{-1.5}\pb$
respectively~\cite{Dittmaier:2011ti,deFlorian:2012yg} (based on
results
including~\cite{Dawson:1990zj,Djouadi:1991tka,Harlander:2002wh,Anastasiou:2002yz,Ravindran:2003um})
and took their 
scale uncertainties to be uncorrelated with those of the efficiencies.
Symmetrising uncertainties, we find correlation coefficients between
the $0$-jet and $\ge 1$-jet cross sections of $-0.43$ ($-0.50$)
for $R=0.4$ ($R=0.5$), using the covariance matrix
in~\cite{BMSZadditional}.

Code to perform the resummations and matchings will be made
available shortly.

This work was supported by the UK STFC, the Agence Nationale de
la Recherche under contract ANR-09-BLAN-0060, the Swiss National Science
Foundation (SNF) under grant 200020-138206 and the European Commission
under contract PITN-GA-2010-264564.
We thank M.~Grazzini and T.~Gehrmann for helpful discussions and
gratefully acknowledge exchanges with T.~Becher and M.~Neubert.

\textbf{Note added:} as our manuscript was being finalised,
Ref.~\cite{Tackmann:2012bt} appeared.
It claims issues in NNLL resummations of jet vetoes, however does not
address the all-order derivation of the NNLL $R$-dependent terms
in~\cite{BSZ12}. Its claim is further challenged by the $\as^3$ numerical
check in Fig.~\ref{fig:checks}.


\newpage

\onecolumngrid
\newpage
\appendix
\section*{Supplemental material}

We here provide material that completes the discussion of the letter,
including more explicit formulae, some derivations and supplementary
figures.

\subsection{Explicit resummation formulae}

In the present section we report the explicit expressions for the resummation functions $g_1$, $g_2$ and $g_3$ computed in 
 \cite{Bozzi:2003jy,Bozzi:2005wk}, as functions of
 $\lambda=\alpha_s\beta_0L$, with $L=\ln(Q/p_t)$. $\alpha_s$ denotes $\alpha_s(\mu_R)$ unless otherwise stated, and $Q$ is the resummation
 scale (see main text)
\begin{subequations}
\begin{align}
  g_{1}(\lambda) &= \frac{A^{(1)}}{\pi\beta_{0}}\frac{2 \lambda +\ln (1-2 \lambda )}{2  \lambda }, \\
  g_{2}(\lambda) &= \frac{1}{2\pi \beta_{0}}\ln (1-2 \lambda )
  \left(A^{(1)} \ln \frac{M^2}{Q^2}+B^{(1)}\right)
  -\frac{A^{(2)}}{4 \pi ^2 \beta_{0}^2}\frac{2 \lambda +(1-2
    \lambda ) \ln (1-2 \lambda )}{1-2
    \lambda} \notag\\
  &+A^{(1)} \bigg(-\frac{\beta_{1}}{4 \pi \beta_{0}^3}\frac{\ln
    (1-2 \lambda ) ((2 \lambda -1) \ln (1-2 \lambda )-2)-4
    \lambda}{1-2 \lambda}-\frac{1}{2 \pi \beta_{0}}\frac{(2 \lambda(1
    -\ln (1-2 \lambda ))+\ln (1-2 \lambda ))}{1-2\lambda} \ln
    \frac{\mu_{R}^2}{Q^2}\bigg)\,,\\
 g_{3}(\lambda) &= \left(A^{(1)} \ln\frac{M^2}{Q^2}+B^{(1)}\right)
    \bigg(-\frac{\lambda }{1-2 \lambda} \ln
   \frac{\mu _{R}^2}{Q^2}+\frac{\beta_{1}}{2 \beta_{0}^2}\frac{2 \lambda
   +\ln (1-2 \lambda )}{1-2 \lambda}\bigg)
   -\frac{1}{2 \pi\beta_{0}}\frac{\lambda}{1-2\lambda}\left(A^{(2)}
       \ln\frac{M^2}{Q^2}+B^{(2)}\right)\notag \\
   &-\frac{A^{(3)}}{4 \pi ^2 \beta_{0}^2}\frac{\lambda ^2}{(1-2\lambda )^2} 
   +A^{(2)} \bigg(\frac{\beta_{1}}{4 \pi  \beta_{0}^3 }\frac{2 \lambda  (3
   \lambda -1)+(4 \lambda -1) \ln (1-2 \lambda )}{(1-2 \lambda
   )^2}-\frac{1}{\pi \beta_{0}}\frac{\lambda ^2 }{(1-2 \lambda )^2}\ln\frac{\mu_{R}^2}{Q^2}\bigg) \notag\\
   & +A^{(1)} \bigg(\frac{\lambda  \left(\beta_{0} \beta_{2} (1-3 \lambda
   )+\beta_{1}^2 \lambda \right)}{\beta_{0}^4 (1-2 \lambda)^2}
   +\frac{(1-2 \lambda) \ln (1-2 \lambda ) \left(\beta_{0} \beta_{2} 
   (1-2 \lambda )+2 \beta_{1}^2 \lambda \right)}{2\beta_{0}^4 (1-2 \lambda)^2} 
   +\frac{\beta_{1}^2}{4 \beta_{0}^4}
   \frac{(1-4 \lambda ) \ln ^2(1-2 \lambda )}{(1-2 \lambda)^2}\notag\\
   &-\frac{\lambda ^2 }{(1-2 \lambda
   )^2} \ln ^2\frac{\mu_{R}^2}{Q^2}
   -\frac{\beta_{1}}{2 \beta_{0}^{2}}\frac{(2 \lambda  (1-2 \lambda)+(1-4 \lambda) \ln (1-2 \lambda ))
   }{(1-2\lambda )^2}\ln\frac{\mu_{R}^2}{Q^2}\bigg),
\end{align}
\end{subequations}
where, for Higgs, $A^{(1)} = 2 C_A$ and $B^{(1)} = -4\pi\beta_0$, while for Drell-Yan,
$ A^{(1)} = 2 C_F$ and $B^{(1)} = -3 C_F$.
 The remaining coefficients can be expressed in a unique way as~\cite{deFlorian:2001zd,Becher:2010tm,Moch:2004pa}:
\begin{align}
     &A^{(2)} = A^{(1)} K_{\rm CMW}^{(1)} ,\quad
     A^{(3)} = A^{(1)} K_{\rm CMW}^{(2)} + \pi\beta_0 C d^{(2)}, \quad
     B^{(2)} = -2\gamma^{(2)}+2 \pi\beta_0 C \zeta_2\,,\\
     &\beta_0 = \frac{11 C_A - 2 n_f}{12\pi}\,,\quad 
     \beta_1 = \frac{17 C_A^2 - 5 C_A n_f - 3 C_F n_f}{24\pi^2}\,,\\ 
     & \beta_2 = \frac{2857 C_A^3+ (54 C_F^2 -615C_F C_A -1415 C_A^2)n_f
       +(66 C_F +79 C_A) n_f^2}{3456\pi^3}\,,\quad 
\end{align}
in terms of the Casimir $C=C_A$ for Higgs and $C=C_F$ for Drell-Yan,
and of the well known constants
\begin{subequations}
\begin{align}
 K_{CMW}^{(1)} &= C_A
 \left(\frac{67}{18}-\frac{\pi^2}{6}\right)-\frac{5}{9}n_f\,,
 \qquad
 d^{(2)} = C_A\left(\frac{808}{27}-28\zeta_{3}\right)-\frac{224}{54}n_f,
 \\
 K_{CMW}^{(2)} &= C_A^2 \left( \frac{245}{24} - \frac{67}{9}\zeta_2
 + \frac{11}{6}\zeta_3 + \frac{11}{5}\zeta_2^2\right) 
+ C_F n_f \left(-\frac{55}{24} + 2\zeta_3\right) 
 + C_A n_f \left(-\frac{209}{108} + \frac{10}{9}\zeta_2 - \frac{7}{3} \zeta_3\right) 
 - \frac{1}{27} n_f^2\,.
\end{align}
\end{subequations}
Here $\gamma^{(2)}$~\cite{Furmanski:1980cm,Curci:1980uw} are the coefficients of the $\delta(1-z)$ term in the NLO splitting 
functions $P^{(1)}$. For Higgs production we have

\begin{equation}
\gamma^{(2)} = C_A^2\left(\frac{8}{3}+3\zeta_3\right)-\frac{1}{2}C_F n_f -\frac{2}{3}C_A n_f\,,  
\end{equation}
whilst for the Drell-Yan process
\begin{equation}
\gamma^{(2)} = C_F^2\left(\frac{3}{8}-\frac{\pi^2}{2}+6\zeta_3\right)+C_FC_A\left(\frac{17}{24}+\frac{11}{18}\pi^2
-3\zeta_3\right)-C_F n_f\left(\frac{1}{12}+\frac{\pi^2}{9}\right)\,.
\end{equation}

We finally report the expressions for the collinear coefficient function $C_{ij}^{(1)}(z)$ 
and the hard virtual term $\mathcal{H}^{(1)}$ in eq.~(\ref{eq:SigmaNNLL-result})
\footnote{Often in the literature, the hard coefficient $H^{(1)}$ is considered as part
 of the $\delta(1-z)$ term in the coefficient function $C_{ij}^{(1)}(z)$, so it comes with a factor
$1/(1-\alpha_s\beta_0L)$ in eq.~(\ref{eq:SigmaNNLL-result}). This results in a different convention for the resummation
coefficient $B^{(2)}$ which will differ by an amount  $2\pi\beta_0H^{(1)}$ from what reported here.}
\begin{subequations}
\begin{align}
 C_{ij}^{(1)}(z) &= -P_{ij}^{(0),\epsilon}(z)-\delta_{ij}\delta(1-z)C \frac{\pi^2}{12}+P_{ij}^{(0)}(z)
  \ln{\frac{Q^2}{\mu_{F}^{2}}}, \\
 \mathcal{H}^{(1)} &= H^{(1)}- \left( B^{(1)}+\frac{A^{(1)}}{2}\ln{\frac{M^2}{Q^2}}\right)\ln{\frac{M^2}{Q^2}}
 + {\rm q}~2\pi\beta_{0}\ln{\frac{\mu_{R}^2}{M^{2}}}\, ,
\end{align}
\end{subequations}
where ${\rm q}$ is the $\alpha_s$ power of the LO cross section (${\rm q} = 2$ for Higgs production and ${\rm q} = 0$ for Drell-Yan).
The coefficient $H^{(1)}$ encodes the pure hard virtual correction to the leading order process, it is given by
\begin{subequations}
\begin{align}
 {\rm Higgs :} \qquad  H^{(1)} &= C_A\left(5+\frac{7}{6}\pi^2\right)-3C_F\,,\\
 {\rm Drell-Yan :} \qquad  H^{(1)} &= C_F\left(-8+\frac{7}{6}\pi^2\right)\,.
\end{align}
\end{subequations}

Finally, $P_{ij}^{(0),\epsilon}(z)$ is the $\mathcal{O}(\epsilon)$ term of the LO splitting function $P_{ij}^{(0)}(z)$:
\begin{subequations}
\begin{align}
 P_{qq}^{(0),\epsilon}(z) &= -C_F(1-z)\,,\\
 P_{gq}^{(0),\epsilon}(z) &= -C_F z\,,\\
 P_{qg}^{(0),\epsilon}(z) &= -z(1-z)\,,\\
 P_{gg}^{(0),\epsilon}(z) &= 0.
\end{align}
\end{subequations}

\subsection{Full matching formulae}
We start by recalling the three prescriptions discussed
in~\cite{BSZ12} for defining the jet-veto efficiency at NNLO
accuracy. In this section, in order to simplify the notation, we will
refer to the integrated jet-veto distribution $\Sigma^{(J)}$ as $\Sigma$. 
\begin{subequations}
\begin{align}
  \label{eq:match_a}
  \epsilon^{(a)}(p_{\rm t, veto}) &= \frac{\Sigma_{0}(p_{\rm t, veto})+\Sigma_{1}(p_{\rm t, veto})+\Sigma_{2}(p_{\rm t, veto})}{\sigma_{0}+\sigma_{1}+\sigma_{2}},\\
 \label{eq:match_b}
  \epsilon^{(b)}(p_{\rm t, veto}) &= \frac{\Sigma_{0}(p_{\rm t, veto})+\Sigma_{1}(p_{\rm t, veto})+\bar{\Sigma}^{(J)}_{2}(p_{\rm t, veto})}{\sigma_{0}+\sigma_{1}},  \\ 
 \label{eq:match_c}
  \epsilon^{(c)}(p_{\rm t, veto}) &= 1+\frac{\bar{\Sigma}^{(J)}_{1}(p_{\rm t, veto})}{\sigma_{0}}
  -\frac{\sigma_{1}}{\sigma_{0}^{2}}\bar{\Sigma}^{(J)}_{1}(p_{\rm t, veto})+\frac{\bar{\Sigma}^{(J)}_{2}(p_{\rm t, veto})}{\sigma_{0}} 
\end{align}
\end{subequations}
with
\begin{align}
 \Sigma_{i}(p_{\rm t, veto}) = \sigma_{i} + \bar{\Sigma}^{(J)}_{i}(p_{\rm t, veto})
\end{align}
being the $\mathcal{O}(\alpha_{s}^{i})-$th correction relative to the Born cross section where
\begin{align}
 \bar{\Sigma}^{(J)}_{i}(p_{\rm t, veto}) = -\int_{p_{\rm t, veto}}^{\infty}~dp_{\rm t}\frac{d\Sigma_{i}(p_{\rm t})}{dp_{\rm t}},
\end{align}
can be determined from MCFM, while $\sigma_i$ is the $i^{\rm th}$
order contribution to the total cross section
(cf.~\cite{Dawson:1990zj,Djouadi:1991tka,Harlander:2002wh,Anastasiou:2002yz,Ravindran:2003um,Hamberg:1990np}).
The above three prescriptions differ by terms $\mathcal{O}(\alpha_{s}^{3})$, which are beyond the control of current fixed-order
calculations.

The jet-veto matched efficiency should tend to one and the differential distribution should vanish at the maximum allowed transverse momentum $p_{\rm t, veto}^{\rm max}$ 
  \begin{align}
   \label{eq:match_conditions}
   \epsilon(p_{\rm t, veto}^{\rm max}) = 1, \qquad \frac{d\epsilon}{dp_{\rm t, veto}}(p_{\rm t, veto}^{\rm max}) = 0. 
  \end{align}
To fulfil such requirements, we modify the resummed logarithms as follows
\begin{align}
 L\rightarrow\tilde{L} = \left(1-\frac{p_{\rm t, veto}}{p_{\rm t, veto}^{\rm max}}\right)
 \frac{1}{p}\ln{\left(\left(\frac{Q}{p_{\rm t, veto}}\right)^{p}-\left(\frac{Q}{p_{\rm t, veto}^{\rm max}}\right)^{p}+1\right)},
\end{align}
where $p$ is some integer power. By default we choose $p=5$~\cite{BSZ12}. 
The factor $\left(1-\frac{p_{\rm t, veto}}{p_{\rm t, veto}^{\rm max}}\right)$ is necessary to 
fulfil eq.~(\ref{eq:match_conditions}) but it is largely irrelevant in practice 
since $p_{\rm t, veto}^{\rm max}$ is much larger than the typical values of the jet transverse momentum
veto (in practice, we set $p_{\rm t, max} = \infty$). We introduce three multiplicative matching schemes~\cite{DISresum}, each of them
corresponding to one of the three efficiency
definitions~(\ref{eq:match_a}), (\ref{eq:match_b}), (\ref{eq:match_c}). 
To simplify the notation, we split the luminosity factor in the square
brackets of Eq.~(\ref{eq:SigmaNNLL-result}) into two terms ${\mathcal
  L}^{(0)}(\tilde{L})$ and ${\mathcal L}^{(1)}(\tilde{L})$, which
start at order $\alpha_s^0$ and $\alpha_s^1$ respectively,
\begin{align}
 \label{eq:L_0}
 {\mathcal L}^{(0)}(\tilde{L}) &= \sum_{i,j}\int dx_1 dx_2 \delta(x_1 x_2 s - M^2)
  f_i\!\left(x_1, e^{-\tilde{L}} \mu_F\right)f_j\!\left(x_2, e^{-\tilde{L}} \mu_F\right), \\ 
 {\mathcal L}^{(1)}(\tilde{L}) &=   \frac{\alpha_{s}}{2\pi}\sum_{i,j}\int dx_1 dx_2  \delta(x_1 x_2 s - M^2)
 \bigg[f_i\!\left(x_1, e^{-\tilde{L}} \mu_F\right)
  f_j\!\left(x_2, e^{-\tilde{L}} \mu_F\right){\cal H}^{(1)} \notag\\
  &+\frac{1}{1-2\alpha_s \beta_0 \tilde{L}}\sum_{k}\bigg(
  \int_{x_1}^1\frac{dz}{z} C_{ki}^{(1)}(z)
  f_i\!\left(\frac{x_1}{z}, e^{-\tilde{L}} \mu_F\right)
  f_j\!\left(x_2, e^{-\tilde{L}} \mu_F\right) + \{(x_1,i)\,\leftrightarrow\,(x_2,j)\}\bigg)\, \bigg].
\end{align}
The first of the three matching schemes then reads
\begin{multline}
\Sigma^{(a)}_{\rm matched}(p_{\rm t, veto}) =\frac{1}{\sigma_0}\frac{\Sigma_{\rm NNLL}(p_{\rm t, veto})}{1+{\mathcal L}^{(1)}(\tilde{L})/{\mathcal L}^{(0)}(\tilde{L})}
 \bigg[\sigma_0\left(1+\frac{{\mathcal L}^{(1)}(\tilde{L})}{{\mathcal L}^{(0)}(\tilde{L})}\right)+\Sigma^{(1)}(p_{\rm t, veto})
 -\Sigma_{\rm NNLL}^{(1)}(p_{\rm t, veto})\\
 +\Sigma^{(2)}(p_{\rm t, veto})-\Sigma_{\rm NNLL}^{(2)}(p_{\rm t, veto})
 +\left(\frac{{\mathcal L}^{(1)}(0)}{{\mathcal L}^{(0)}(0)}
 -\frac{\Sigma_{\rm NNLL}^{(1)}(p_{\rm t, veto})}{\sigma_{0}}\right)
 \left(\Sigma^{(1)}(p_{\rm t, veto})-\Sigma_{\rm NNLL}^{(1)}(p_{\rm t, veto})\right)\bigg],
\end{multline}
and the corresponding jet-veto efficiency is
\begin{equation}
 \epsilon^{(a)}_{\rm matched}(p_{\rm t, veto}) = \frac{\Sigma^{(a)}_{\rm matched}(p_{\rm t, veto})}{\Sigma^{(a)}_{\rm matched}(p_{\rm t, veto}^{\rm max})}\,.
\end{equation}
The second scheme can be derived from the previous one by replacing
$\Sigma^{(2)}(p_{\rm t, veto})$ with  $\bar{\Sigma}^{(2)}(p_{\rm t, veto})$. For the 
vetoed cross section we get
\begin{multline}
 \Sigma^{(b)}_{\rm matched}(p_{\rm t, veto}) =\frac{1}{\sigma_0}\frac{\Sigma_{\rm NNLL}(p_{\rm t, veto})}{1+{\mathcal L}^{(1)}(\tilde{L})/{\mathcal L}^{(0)}(\tilde{L})}
 \bigg[\sigma_0\left(1+\frac{{\mathcal L}^{(1)}(\tilde{L})}{{\mathcal L}^{(0)}(\tilde{L})}\right)+\Sigma^{(1)}(p_{\rm t, veto})
 -\Sigma_{\rm NNLL}^{(1)}(p_{\rm t, veto})\\
 +\bar{\Sigma}^{(2)}(p_{\rm t, veto})-\Sigma_{\rm NNLL}^{(2)}(p_{\rm t, veto})
 +\left(\frac{{\mathcal L}^{(1)}(0)}{{\mathcal L}^{(0)}(0)}
 -\frac{\Sigma_{\rm NNLL}^{(1)}(p_{\rm t, veto})}{\sigma_{0}}\right)
 \left(\Sigma^{(1)}(p_{\rm t, veto})-\Sigma_{\rm NNLL}^{(1)}(p_{\rm t, veto})\right)\bigg],
\end{multline}
while for its efficiency
\begin{equation}
 \epsilon^{(b)}_{\rm matched}(p_{\rm t, veto}) = \frac{\Sigma^{(b)}_{\rm matched}(p_{\rm t, veto})}{\Sigma^{(b)}_{\rm matched}(p_{\rm t, veto}^{\rm max})}\,.
\end{equation}
Finally, the third matching scheme is directly formulated for the efficiency resulting in 
\begin{multline}
 \epsilon^{(c)}_{\rm matched}(p_{\rm t, veto}) =\frac{1}{\sigma_0^2}\frac{\Sigma_{\rm NNLL}(p_{\rm t, veto})}{1+{\mathcal L}^{(1)}(\tilde{L})/{\mathcal L}^{(0)}(\tilde{L})} 
 \bigg[\sigma_0\left(1+\frac{{\mathcal L}^{(1)}(\tilde{L})}{{\mathcal L}^{(0)}(\tilde{L})}\right)
  +\bar{\Sigma}^{(1)}(p_{\rm t, veto})
 -\Sigma_{\rm NNLL}^{(1)}(p_{\rm t, veto})\\
 +\bar{\Sigma}^{(2)}(p_{\rm t, veto})-\frac{\sigma_1}{\sigma_0}\bar{\Sigma}^{(1)}(p_{\rm t, veto})-\Sigma_{\rm NNLL}^{(2)}(p_{\rm t, veto})\\
 +\left(\frac{{\mathcal L}^{(1)}(0)}{{\mathcal L}^{(0)}(0)}
 -\frac{\Sigma_{\rm NNLL}^{(1)}(p_{\rm t, veto})}{\sigma_{0}}\right)
 \left(\bar{\Sigma}^{(1)}(p_{\rm t, veto})-\Sigma_{\rm NNLL}^{(1)}(p_{\rm t, veto})\right)\bigg].
\end{multline}

\subsection{Details of relation between jet and boson-$p_t$ resummations}

This section collects a number of results to help relate jet and boson
$p_t$ resummations. Firstly we demonstrate that the $g_n(\as L)$ from
a boson $p_t$ resummation can be directly carried over to jet $p_t$
resummation for $n \le 3$.
Then we obtain a form for the boson $p_t$ resummation that is suitable
for expansion and comparison with fixed-order results.
Finally we consider the large $R$ limit of the jet-$p_t$ resummation,
which was used in~\cite{Becher:2012qa} to attempt to obtain a relation
between jet and boson $p_t$ resummations.

\subsubsection{Relating $g_n(\as L)$ between boson and jet resummations}

One of the main ingredients of our results are the $g_n(\as L)$
functions that are used in boson $p_t$ resummations. These stem
from the rightmost integral in Eq.~(\ref{eq:ptB}), which involves a
Fourier transformation, whereas in Eq.~(\ref{eq:ptJ-start}) we need
related integrals but with a theta-function instead of the
$(\exp(ib.k_t)-1)$ factor.

We start from the expression for the resummed $p_t$ distribution in
eq.~\eqref{eq:ptB} and concentrate on the part of the matrix-element
in the right-hand integral that is responsible for the leading
logarithms.
Integrating over azimuthal angles we obtain:
\begin{equation}
  \label{eq:ptBresum}
  \frac{d\Sigma^{(B)}(p_t)}{p_t dp_t} = \sigma_0 \int b db J_0(b p_t)
  \exp [-{\cal R}(b)]\,,
  \qquad
  {\cal R}(b)=\int [dk] M^2(k)  (1-J_0(b k_t))\,.
\end{equation}
%
We wish to show that we can safely perform the replacement
\begin{equation}
  \label{eq:J0theta}
  (1-J_0(b k_t)) \to \Theta(k_t-b_0/b)\,,\quad
  b_0 = 2 e^{-\gamma_E}\,,
\end{equation}
up to and including NNLL accuracy.
Integrating over rapidity ${\cal R}(b)$ has the form
\begin{equation}
  \label{eq:ptRsc}
  {\cal R}(b)=\int_0^{M} \frac{dk_t}{k_t} F\left(\alpha_s \ln \frac{M}{k_t}\right)  (1-J_0(b k_t))\,,
\qquad
F\left(\alpha_s \ln \frac{M}{k_t}\right) = 4 C\frac{\alpha_s}{\pi}\ln\frac{M}{k_t}\frac{1}{1-2 \alpha_s\beta_0 \ln\frac{M}{k_t}}\,.
\end{equation}
To evaluate separately real and virtual contributions in eq.~\eqref{eq:ptRsc}, we introduce a dimensional regulator and write
\begin{equation}
  \label{eq:ptRsc-reg}
  {\cal R}(b)=F\left(\alpha_s \partial_\epsilon\right) \left.\int_0^{M} \frac{dk_t}{k_t} 
  \left(\frac{k_t}{M}\right)^{-\epsilon}  (1-J_0(b k_t))\right|_{\epsilon=0}\,,
\end{equation}
which yields
\begin{equation}
  {\cal R}(b)=R_{\rm LL}(b_0/b)+ \delta {\cal R}(b)\,, 
\end{equation}
where, neglecting terms suppressed by powers of $1/(b M)$, 
\begin{equation}
  \label{eq:Rb}
  R_{\rm LL}(b_0/b)=\int_0^{M} \frac{dk_t}{k_t} F\left(\alpha_s \ln \frac{M}{k_t}\right) \Theta(k_t-b_0/b)\,,
\end{equation}
and
\begin{subequations}
  \label{eq:deltaR}
\begin{align}
  \delta {\cal R}(b)&= F\left(\alpha_s \partial_\epsilon\right)
  \left.\frac{(b/b_0)^\epsilon}{\epsilon}\left[-1+e^{-\gamma_E \epsilon}\frac{\Gamma(1-\frac{\epsilon}{2})}{\Gamma(1+\frac{\epsilon}{2})}\right]\right|_{\epsilon=0}\\
&=F\left(\alpha_s \partial_\epsilon\right) \left.\left(\frac{b}{b_0}\right)^\epsilon\left[\frac{\zeta_3}{12} \epsilon^2+{\cal O}(\epsilon^4)\right]\right|_{\epsilon=0}\,.
\end{align}
\end{subequations}
This gives at most a term $\alpha_s^n \ln^{n-2}(M b/b_0)$,
i.e.~a N$^3$LL term. 
A similar argument can be applied to contributions to ${\cal R}(b)$ arising
from less singular regions, giving also rise to terms that are beyond 
NNLL.
Consequently, to NNLL accuracy, the same $g_1$, $g_2$ and $g_3$
functions can be used in both the jet and boson resummation.

\subsubsection{Evaluation of the boson-$p_t$ integrated cross section}

To facilitate comparisons between the jet and boson $p_t$ resummations
at fixed order, it is convenient to have an expression for the boson
$p_t$ resummation whose fixed-order expansion can be straightforwardly
obtained. 
The full expression for the cumulative $p_t$ cross section can be
found in~\cite{Bozzi:2003jy,Bozzi:2005wk,Becher:2010tm} and reads
\begin{align}
  \label{eq:ptBsigma}
  \Sigma^{(B)}(p_t) &=
  \int_0^{\infty} dy J_1(y) \,|{\cal M}_B|^2 e^{-R(b_0/b)}
  \left({\cal L}^{(0)}(\ln(Qb/b_0))+{\cal L}^{(1)}(\ln(Qb/b_0))\right)\,,
\end{align}
where
\begin{equation}
  -R(b_0/b) =  \ln(Qb/b_0) g_1(\as \ln(Qb/b_0)) + g_2(\as \ln(Qb/b_0)) +
  \frac{\as}{\pi} g_3(\as \ln(Qb/b_0))
\end{equation}
is the full NNLL radiator. As discussed above, the
resummation functions $g_1$, $g_2$ and $g_3$ are those used for the
jet veto case.  To perform the inverse Fourier transform we expand
$R(b_0/b)$ and the full luminosity factor around $b=b_0/p_t$ and
neglect subleading logarithmic terms getting, at NNLL accuracy,
\begin{align}
  \label{eq:pt-cross-section-result}
  \Sigma^{(B)}(p_t) =
  \int_0^{\infty} dy J_1(y) \,|{\cal M}_B|^2&\bigg[{\cal L}^{(0)}(\ln(Q/p_t))+{\cal L}^{(1)}(\ln(Q/p_t))
  +\partial_{\ln p_t}{\cal L}^{(0)}(\ln(Q/p_t))\ln(y/b_0)\bigg]\notag\\
  &\times\left(\frac{y}{b_0}\right)^{-R'}e^{-R(p_t)}
  \left(1
    -\frac{1}{2}R''\ln^2(y/b_0)\right)\,,
\end{align}
where we have performed the change of variable $y = b p_t$, and we
have made use of $R'$ and $R''$, the first and second derivatives of
$R$ with respect to $\ln(Q/p_t)$.
To order $\as L$, $R' = 4 \as C \ln (Q/p_t) /\pi$.
Moreover, from eq.~(\ref{eq:L_0}), we see that the variation of ${\cal L}^{(0)}(L)$ reads
\begin{align}
 \partial_{\ln p_t}{\cal L}^{(0)}(L) = \frac{\alpha_s}{\pi}\sum_{i,j,k} \int dx_1 dx_2 
 \delta(x_1 x_2 s - M^2)\left[(P_{ki}^{(0)}\otimes f_i)\left(x_1, e^{-L} \mu_F\right)
 f_j\!\left(x_2, e^{-L} \mu_F\right)
 + \{(x_1,i)\,\leftrightarrow\,(x_2,j)\}\right].
\end{align}
It is straightforward to show that eq.~(\ref{eq:pt-cross-section-result}) evaluates to
\begin{multline}
  \label{eq:pt-kernel}
  \Sigma^{(B)}(p_t) =
  \,|{\cal M}_B|^2e^{-R(p_t)}\bigg[{\cal L}^{(0)}(\ln(Q/p_t))
  \left(1
    -\frac{1}{2}R''\partial_{R'}^2\right)\\
  +{\cal L}^{(1)}(\ln(Q/p_t))
  -\partial_{\ln p_t}{\cal L}^{(0)}(\ln(Q/p_t))\partial_{R'}\bigg]
  e^{-\gamma_E R'}\frac{\Gamma(1-\frac{R'}{2})}{\Gamma(1+\frac{R'}{2})}\,.
\end{multline}
In this notation, the result for the jet-veto cross section is simply
$|{\cal M}_B|^2e^{-R(p_t)}({\cal L}^{(0)} +  {\cal L}^{(1)}) (1 +
\cF^{\text{clust}} + \cF^{\text{correl}})$.
It is therefore immediate to evaluate the differences between the two
formulae at any given fixed order and in particular to derive
Eq.~(\ref{eq:D2Diff}): making use of the fact that $e^{-\gamma_E
  R'}{\Gamma(1-\frac{R'}{2})}/{\Gamma(1+\frac{R'}{2})}$ has an
expansion of the form $1 + \frac{\zeta_3}{12}{R'}^3 + \order{{R'}^5}$,
one sees that the only terms in the difference that survive at order
$\as^2 L$ are the ${\cal F}^\text{clust}$ and ${\cal F}^\text{correl}$
contributions and the $R'' \partial_{R'}^2$ term of
Eq.~(\ref{eq:pt-kernel}), with the latter giving
\begin{equation}
\label{eq:expansion-pt-kernel}
   -\sigma_0 \frac{1}{2}R''\partial_{R'}^2
  e^{-\gamma_E R'}\frac{\Gamma(1-\frac{R'}{2})}{\Gamma(1+\frac{R'}{2})}
  = \sigma_0\left(-4\frac{\alpha_s^2}{\pi^2}\zeta_3 C^2
    \ln\frac{Q}{p_t}+ \order{\as^2 L^0} + \order{\as^3 L^2} 
  \right)\,,
\end{equation}
which is the source of the $\zeta_3$ in Eq.~(\ref{eq:D2Diff}).\footnote{
One point to note in evaluating the difference between the jet and
boson $p_t$
resummations at order $\as^3 L^2$ is that it is necessary to account
also for the difference between $C_2$ terms for the two resummations.
One of the properties of this difference of $C_2$ terms is that is has
$Q$ dependence that ensures that the final prediction for the
difference of $\as^3 L^2$ terms is $Q$-independent.
To produce figure~\ref{fig:checks} the difference of $C_2$
terms was taken from a numerical determination based on the MCFM
leading-order $H+2$-jet calculation.
}

\subsubsection{Use of the large-$R$ limit to relate boson and
  jet-$p_t$ resummations}

One natural way of relating jet and boson-$p_t$ resummations is to make
the observation that for an infinite jet radius, all partons will be
clustered into a single jet, which will have a transverse momentum
that balances exactly that of the boson.
This approach was taken in Ref.~\cite{Becher:2012qa} and here we
examine it in detail.

First, let us consider the properties of $\cF^{\text{clust}}$ and 
$\cF^{\text{correl}}$ for large $R$. 
It is straightforward to see that $\cF^\text{correl}$ vanishes for
large $R$, since in Eq.~(\ref{eq:Fcorrel}) the two partons will always
clustered together, giving $1-J(k_1,k_2) = 0$.
For $\cF^\text{clust}$, the NNLL component for $R>\pi$ can be
evaluated in closed form and is given by
\begin{equation}
{\cal F}^\text{clust} = - 4 \frac{\as^2(\ptjv) C^2}{\pi^2} \ln (Q/p_t)
\bigg(\left(\frac{\pi}{6}R^{2}-\frac{R^{4}}{8\pi}\right)\arctan\frac{\pi}{\sqrt{R^2-\pi^2}}
 +\left(\frac{R^2}{8}-\frac{\pi^2}{12}\right)\sqrt{R^2-\pi^2}\bigg).
\end{equation}
This has the property that it vanishes as $1/R$ for large $R$.
Thus it would appear that at order $\as^2 L$ the difference between
jet and boson-$p_t$ resummations should be given
by~\cite{Becher:2012qa} 
\begin{equation}
  \label{eq:D2Diff-BN}
  \frac{d\Sigma^{(J)}_{\NNLL,2}(p_t)}{d\ln p_t} -
  \frac{d\Sigma^{(B)}_{\NNLL,2}(p_{t})}{d\ln p_{t}} = 
  (f(R) - f(\infty)) \, \as^2 \sigma_0 = f(R) \, \as^2 \sigma_0\,,
\end{equation}
which differs from the result in Eq.~(\ref{eq:D2Diff}) (here
$f(R) = f^\text{correl}(R) + f^\text{clust}(R)$).

To understand the origin of this difference, it is helpful to examine the
structures that lead to ${\cal F}^\text{clust}$ vanishing for large
$R$.
A first observation is that for large $R$, $J(k_1,k_2)$ can be
written as
\begin{equation}
  \label{eq:J-large-R}
  J(k_1, k_2) 
  = \Theta\left(R - |\Delta y| + \frac{\Delta\phi^2}{2R} 
               + \order{\frac{1}{R^3}}\right)
  \,,
  \qquad
  \Delta y \equiv y_1 - y_2\,,\quad \Delta \phi = \phi_1 - \phi_2\,.
\end{equation}
Neglecting the term of order $1/R$ will allow us to simplify our
discussion and so we will instead examine a ``rapidity-only'' jet
algorithm with the clustering function 
\begin{equation}
  \label{eq:J-rap}
  J_\text{rap}(k_1, k_2) = \Theta(R - |\Delta y|)\,.
\end{equation}
Let us now evaluate Eq.~(\ref{eq:ptJ-clustering}) with
$J_\text{rap}$. 
We break the problem into rapidity, transverse momentum
and azimuthal integrals.
Each emission $i$ is limited to a rapidity $|y_i| < \ln(M/k_{ti})$. Assuming that we can neglect terms $\ln (k_{t1}/k_{t2})$ from
the rapidity integration, we can write the latter as
\begin{equation}
  \label{eq:rap-integrals}
  \int dy_1 dy_2 
  \Theta\left(|y_1| - \ln \frac{M}{k_{t1}}\right) 
  \Theta\left(|y_2| - \ln \frac{M}{k_{t2}}\right)
  \Theta(R - |y_1 - y_2|)
  = 4R \ln \frac{M}{k_{t1}} - R^2 + \order{R \ln \zeta}\,,
\end{equation}
where $\zeta = k_{t2}/k_{t1}$ and we have included the constraint that
$J_\text{rap}(k_1,k_2)$ be non-zero.
We can then write Eq.~(\ref{eq:ptJ-clustering}) as
\begin{equation}
  \label{eq:Fclust-large-R}
 {\cal F}^{\rm clust} = 
 4\frac{\alpha_s^2 C^2}{\pi^2}
 \int_0^1\frac{d\zeta}{\zeta}
 \int_{-\pi}^{\pi}\frac{d\phi}{2\pi}
 \int_{p_t}^{\frac{p_t}{\sqrt{1+\zeta^2+2\zeta\cos\phi}}}\frac{dk_{t,1}}{k_{t,1}}
 \left(4R \ln \frac{M}{k_{t1}} - R^2\right)\,,
\end{equation}
where we have dropped the $\order{R \ln\zeta}$ term of
Eq.~(\ref{eq:rap-integrals}).
Performing the $k_{t1}$ integration gives
\begin{equation}
  \label{eq:Fclust-large-R-ptint-done}
 {\cal F}^{\rm clust} = 
 4\frac{\alpha_s^2 C^2}{\pi^2}
 \int_0^1\frac{d\zeta}{\zeta}
 \int_{-\pi}^{\pi}\frac{d\phi}{2\pi}
 \left[\left((-2R \ln \frac{M}{p_t} + \frac{R^2}2 \right)
               \ln (1+\zeta^2 + 2\zeta\cos\phi) 
     - \frac{R}2 \ln^2 (1+\zeta^2 + 2\zeta\cos\phi)
   \right]\,.
\end{equation}
Because $\int_0^{2\pi} d\phi \ln (1+\zeta^2 + 2\zeta\cos\phi) = 0$,
the first term in square brackets vanishes.
This was the only term that had a NNLL $\as^2 \ln M/p_t$ factor and so
at NNLL accuracy ${\cal F}^{\rm clust}$ is zero at large $R$, modulo
$1/R$ corrections associated with the $1/R$ term in
Eq.~(\ref{eq:J-large-R}).
The only element that survives the azimuthal integration in
Eq.~(\ref{eq:Fclust-large-R-ptint-done}) is the second term in square
brackets, resulting in
\begin{equation}
  \label{eq:Fclust-large-R-result}
 {\cal F}^{\rm clust} = -2\frac{\alpha_s^2 C^2}{\pi^2} R\, \zeta_3\,.
\end{equation}
This is N$^3$LL, so beyond our accuracy.
Note, however, that it is enhanced by a factor of $R$.
In the large $R$ limit, the separation between partons is limited to
be at most $2\ln M/p_t$ and thus the $R$ factor is effectively
replaced with a coefficient of order $\ln M/p_t$.
Consequently the apparently N$^3$LL term of
Eq.~(\ref{eq:Fclust-large-R-result}) is ``promoted'' and becomes a
NNLL $\as^2 \ln M/p_t$ contribution.
This is not accounted for in the purely NNLL $R$-dependent analysis that
led to Eq.~(\ref{eq:D2Diff-BN}).

The exact infinite $R$ result can be obtained at order $\as^2 L$ by
evaluating $\cF^\text{clust}$ with $J(k_1,k_2) = 1$, giving
\begin{equation}
 {\cal F}^{\rm clust} = 16\frac{\alpha_s^2 C^2}{\pi^2}\int_0^1\frac{d\zeta}{\zeta}\int_{-\pi}^{\pi}\frac{d\phi}{2\pi}
  \int_{p_t}^{\frac{p_t}{\sqrt{1+\zeta^2+2\zeta\cos\phi}}}\frac{dk_{t,1}}{k_{t,1}}\left(\ln\frac{M}{k_{t,1}}-\ln\zeta\right)\ln\frac{M}{k_{t,1}}
  =-4\frac{\alpha_s^2}{\pi^2}\zeta_3 C^2 \ln\frac{M}{p_t}+\mathcal{O}(\alpha_s^2 \ln^0\frac{M}{p_t}).
\end{equation}
Note the agreement of the $\zeta_3$ term here with that derived in
Eq.~(\ref{eq:expansion-pt-kernel}).
It is this contribution that corresponds to the $\zeta_3$ term in
Eq.~(\ref{eq:D2Diff}).

\subsection{Correlation matrix between 0-jet and  inclusive 1-jet cross
sections} 

As discussed in \cite{BSZ12}, the prescription that we propose for
determining the uncertainties on the 0-jet cross section
 is to treat the uncertainties on the jet-veto
efficiency and on the total cross section as uncorrelated.
This gives the following covariance matrix for the uncertainties of
the 0-jet ($\sigma_\text{0-jet}$) and inclusive 1-jet ($\sigma_{\ge
  \text{1-jet}}$) cross sections:
\begin{equation}
  \label{eq:covariance-matrix}
  \left(
    \begin{array}{cc}
    \epsilon^2 \delta_{\sigma}^2 + \sigma^2 \delta_\epsilon^2
    &
    \epsilon(1-\epsilon)\delta_{\sigma}^2 - \sigma^2 \delta_\epsilon^2
    \\
    \epsilon(1-\epsilon)\delta_{\sigma}^2 - \sigma^2 \delta_\epsilon^2
    &
    (1-\epsilon)^2 \delta_{\sigma}^2 + \sigma^2 \delta_\epsilon^2
    \end{array}
  \right)
\end{equation}
\subsection{Results at 7 TeV}

\begin{figure*}[t]
  \centering
  \includegraphics[width=0.48\linewidth]{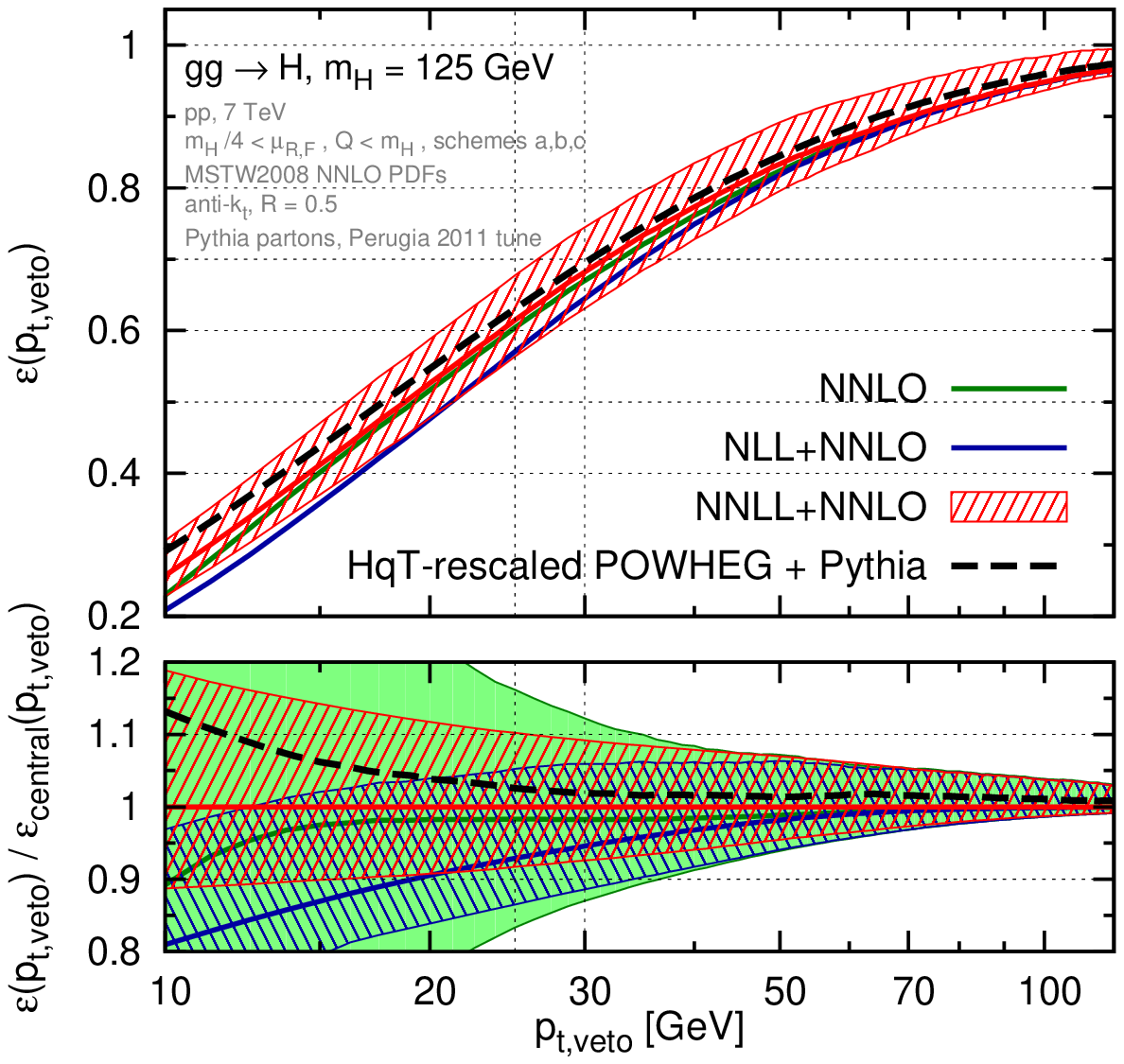}\hfill
  \includegraphics[width=0.49\linewidth]{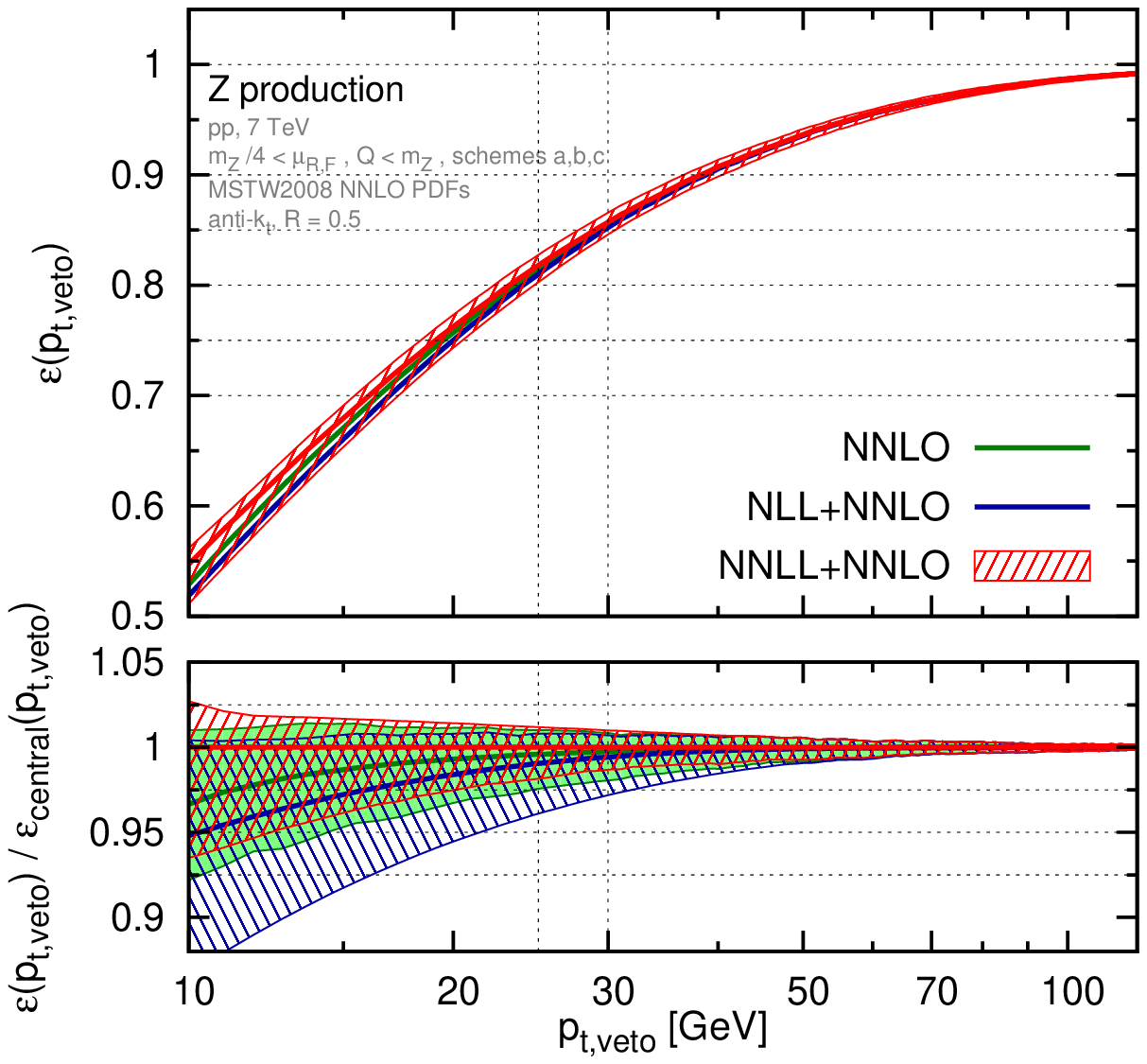}
  \caption{ %
    Comparison of NNLO, NLL+NNLO and NNLL+NNLO results for the
    jet-veto efficiency for Higgs  (left) and Z-boson (right)
    production at 7 TeV.  %
    The Higgs plot also includes the result from a POWHEG (revision
    1683) \cite{Nason:2004rx,Alioli:2008tz} plus Pythia (6.426)
    \cite{Sjostrand:2006za,Skands:2010ak} simulation in which the Higgs-boson $p_t$
    distribution has been reweighted to match the NNLL+NNLO prediction
    from HqT~2.0~\cite{Bozzi:2005wk} as in \cite{BSZ12}.
    The lower panels show the results normalised to the central
    NNLL+NNLO curves.}
  \label{fig:two-matched-v-rwgtPWG-LHC7}
\end{figure*}

For completeness, we show in Fig.~\ref{fig:two-matched-v-rwgtPWG-LHC7}
results for $7\TeV$ centre of mass energy. 
The changes relative to the $8\TeV$ results are modest, with very
slightly higher efficiencies at $7\TeV$.
This can be understood because at higher centre of mass energy, the
PDFs are probed at lower $x$ values, where the scale dependence is
steeper, causing the efficiencies to drop off more rapidly as one
decreases $\ptjv$.

\subsection{R dependence of results}

\begin{figure}
  \centering
  \includegraphics[width=0.48\linewidth]{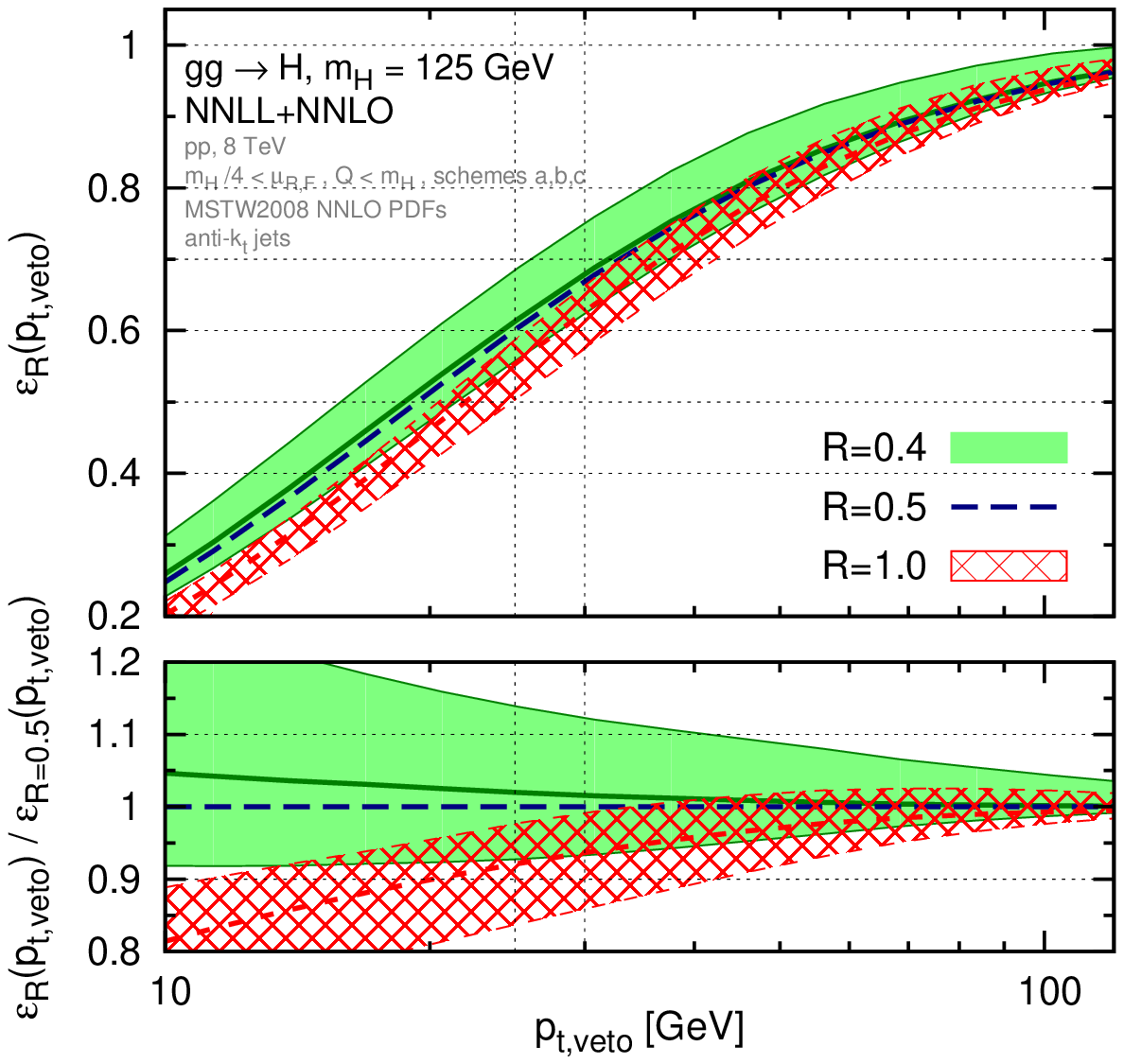}\hfill
  \includegraphics[width=0.49\linewidth]{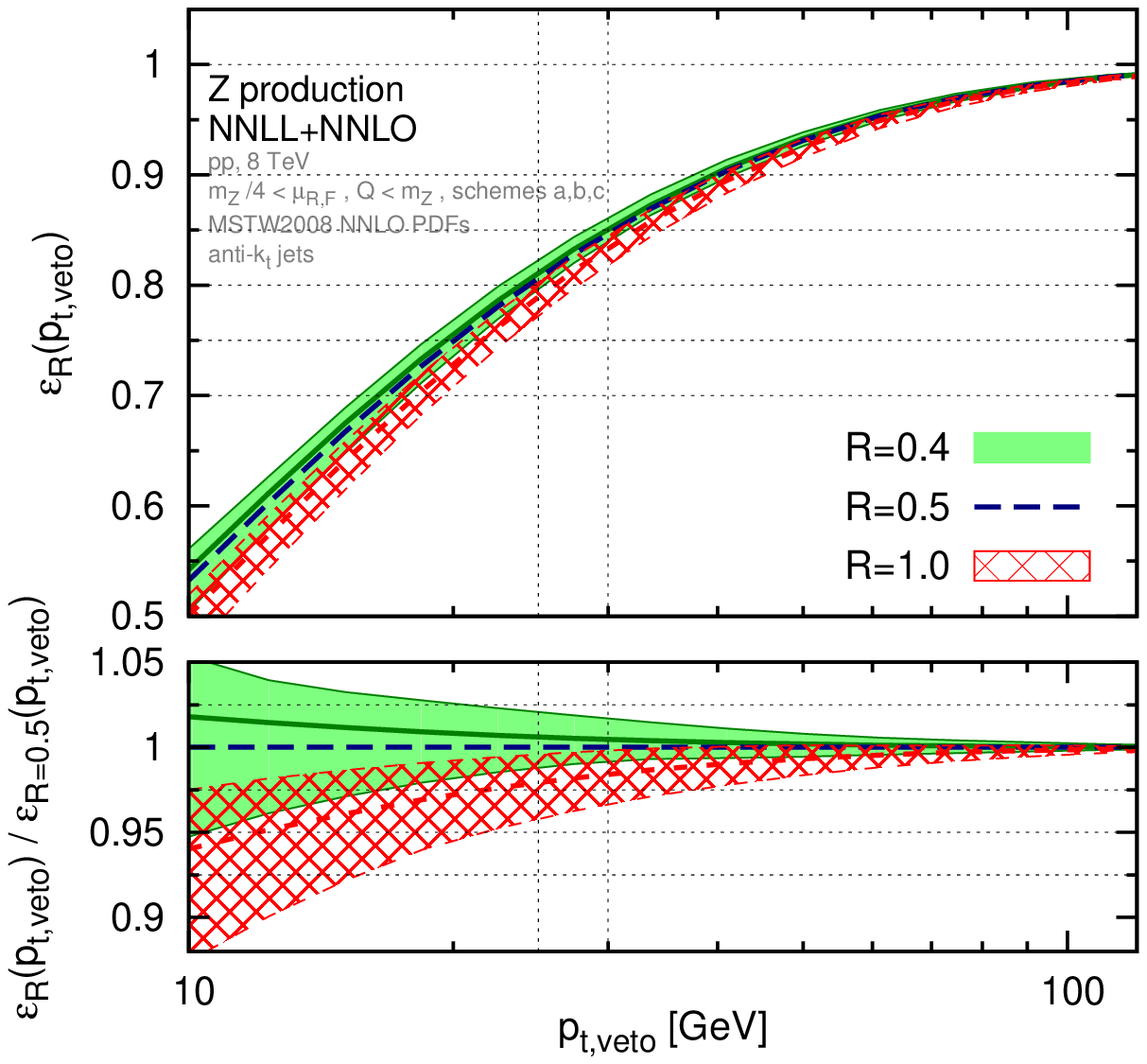}\hfill
  \caption{Jet veto efficiency at NNLL+NNLO as a function of $\ptjv$,
    comparing several jet-radius values; shown for $pp$ collisions at
    a centre-of-mass energy of $8\TeV$, for gluon-fusion Higgs
    production with $M_H = 125\GeV$ (large $m_{\text{top}}$ limit) and
    for $Z$-boson production.
    Uncertainty bands are shown only for $R=0.4$ and $R=1.0$ in order
    to enhance the clarity of the figure. The $R=0.5$ uncertainty band
    is to be found in Fig.~\ref{fig:two-matched-v-rwgtPWG}.
    The lower panels show the predictions normalised to the central
    $R=0.5$ results.  }
  \label{fig:threeR}
\end{figure}

Figure~\ref{fig:threeR} shows the the jet veto efficiency as a
function of $\ptjv$ for several different jet-radius ($R$) values. 
Increasing the jet radius, more radiation is captured and therefore a
jet is more likely to pass the $\ptjv$ threshold and so be vetoed.
Consequence the jet-veto efficiency is expected to be lower for larger
$R$ values. 
This is precisely as observed in Fig.~\ref{fig:threeR}.

Quantitatively, the differences between the $R=0.4$ and $R=0.5$
results (the values used respectively by ATLAS and CMS) are small
compared to the uncertainties on the predictions.
In contrast, for $R=1$ the differences compared to the smaller-$R$
results are not negligible.
One interesting feature, commented on briefly in the main text, is
that for the Higgs-boson case, the uncertainties are somewhat smaller
for $R=1$ than for $R=0.4$ and $R=0.5$, especially the upper part of
the uncertainty band.
This can be understood with the help of the observation that the upper
edge of the uncertainty band for the small $R$ values is set by the
$Q=M_H$ variant of the resummation (recall that our default $Q$ is
$M_H/2$).
Using $Q = M_H$ increases the size of $L$.
Since the $f^{\text{correl}}(R)+f^\text{clust}(R)$ function grows
for small $R$ and multiplies $\as^2 L$, a smaller $R$ value
magnifies the impact of an increase in $Q$.

If, experimentally, one were to consider using larger $R$ values for
performing jet vetoes in order to reduce the theoretical
uncertainties, one concern might be the greater contamination of the
jet's $p_t$ from the underlying event and pileup.
To some extent this could be mitigated by methods such as
subtraction~\cite{Cacciari:2007fd},
filtering~\cite{Butterworth:2008iy} or trimming~\cite{Krohn:2009th}.
Note that with subtraction and filtering (when the latter uses
two filtering subjets, or more) our jet-veto predictions remain
unchanged at NNLO and at NNLL accuracy.

\end{document}